\documentclass[nohyperref]{article}

\usepackage{microtype}
\usepackage{graphicx}
\usepackage{subfigure}
\usepackage{booktabs} 

\usepackage{hyperref}

\usepackage[accepted]{icml2023}

\usepackage[normalem]{ulem}

\usepackage{amsmath}
\usepackage{amssymb}
\usepackage{mathtools}
\usepackage{amsthm}

\usepackage[capitalize,noabbrev]{cleveref}

\theoremstyle{plain}

\theoremstyle{definition}

\theoremstyle{remark}

\usepackage[textsize=tiny]{todonotes}

\icmltitlerunning{AbMPNN ICML-WCB 2023}

\begin{document}

\twocolumn[
\icmltitle{Inverse folding for antibody sequence design using deep learning}


\icmlsetsymbol{equal}{*}

\begin{icmlauthorlist}
\icmlauthor{Frédéric A. Dreyer}{equal,exs}
\icmlauthor{Daniel Cutting}{equal,exs}
\icmlauthor{Constantin Schneider}{exs}
\icmlauthor{Henry Kenlay}{exs}
\icmlauthor{Charlotte M. Deane}{exs}
\end{icmlauthorlist}

\icmlaffiliation{exs}{Exscientia, Oxford Science Park, Oxford, OX4 4GE, UK}

\icmlcorrespondingauthor{Frédéric A. Dreyer}{fdreyer@exscientia.co.uk}

\icmlkeywords{Machine Learning, ICML}

\vskip 0.3in
]



\printAffiliationsAndNotice{\icmlEqualContribution} 

\begin{abstract}
We consider the problem of antibody sequence design given 3D structural information. 
Building on previous work, we propose a fine-tuned inverse folding model that is specifically optimised for antibody structures and outperforms generic protein models on sequence recovery and structure robustness when applied on antibodies, with notable improvement on the hypervariable CDR-H3 loop. 
We study the canonical conformations of complementarity-determining regions and find improved encoding of these loops into known clusters.
Finally, we consider the applications of our model to drug discovery and binder design and evaluate the quality of proposed sequences using physics-based methods.

\end{abstract}

\section{Introduction}
\label{submission}

In recent years there has been rapid progress in tackling the problem of protein folding: predicting the three-dimensional structure of a protein based only on its sequence data. Machine learning methods 
have been shown to achieve new standards of accuracy, often predicting complicated structures with experimental level accuracy e.g.~\cite{af2,rosettafold,esmfold}.

\textit{De novo} protein design can be described as an inverse folding problem: given a backbone structure with atomic coordinates, what amino acid sequences will fold to this shape? Solving this problem has implications in a range of important applications in protein engineering, from the design of novel enzymes, receptors and other biomolecules with tailored functions, to drug discovery for efficiently exploring binder designs that can target a specific protein's active site. It is also of high relevance in applications of certain machine learning models, such as generative diffusion models. Residues are then often represented as non-specific backbone nodes in the form of $C_\alpha$ atom coordinates and frame orientations such that a sequence needs to be designed from the predicted structure~\cite{rfdiffusion,genie,framediff}.

Inverse folding has historically been approached as an energy optimisation problem, using tools such as Rosetta to search for combinations of amino acid identities and conformations that result in the lowest energy for a given structure~\cite{rosetta}. Recent advances in deep learning have offered an alternative data-driven approach which results in significantly faster and often more accurate models~\cite{structransfo,STROKACH2020402,AnandAchim,jing2021learning,esmif}. 

\begin{figure}
    \centering
    \includegraphics[width=0.65\linewidth]{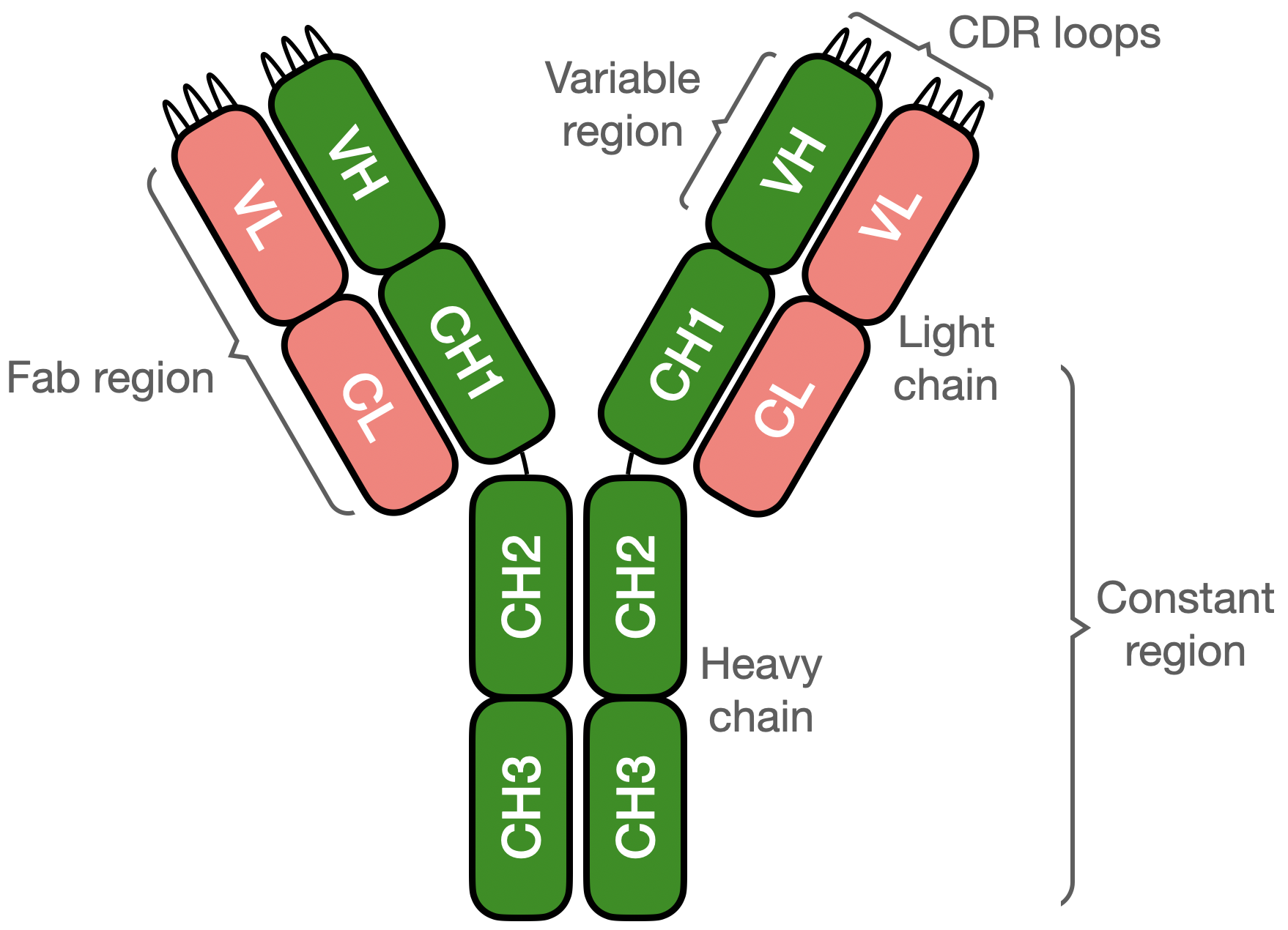}
    \caption{Overview of an antibody structure and its domains.}
    \label{fig:ab}
\end{figure}

In this study, we consider the structured graph neural network method used in the recent ProteinMPNN model~\cite{proteinmpnn} to build an antibody-specific inverse folding model. Antibodies, illustrated in Figure~\ref{fig:ab}, are proteins that play a central role in the adaptive immune system due to their ability to bind to a wide range of pathogens. They consist of two heavy and two light chains divided into domains of approximately 110 amino acid residues, with the N-terminal domains, called variable regions, each containing three hypervariable loops known as the complementarity determining regions (CDRs) that make up the majority of the antigen binding~\cite{cdrs}.
We create our model by fine-tuning on data from the Structural Antibody Database (SAbDab)~\cite{sabdab1,sabdab2}, as well as on paired sequences from the Observed Antibody Space (OAS)~\cite{oas1,oas2} using structures predicted by the ABodyBuilder2 model~\cite{abb2}. We show that our approach provides state-of-the-art performance in predicting the amino acid residues in the CDR loops, with notable improvement in sequence recovery and designability for the third CDR loop of the heavy chain (CDR-H3), the most sequentially and structurally diverse loop and typically the most important region for antigen recognition~\cite{cdrh3_2011,cdrh3}.
We publish our model weights to allow further downstream applications~\cite{frederic_a_dreyer_2023_8164693}.

\section{Inverse folding with ProteinMPNN}
\begin{figure}
    \centering
    \includegraphics[width=1.0\linewidth]{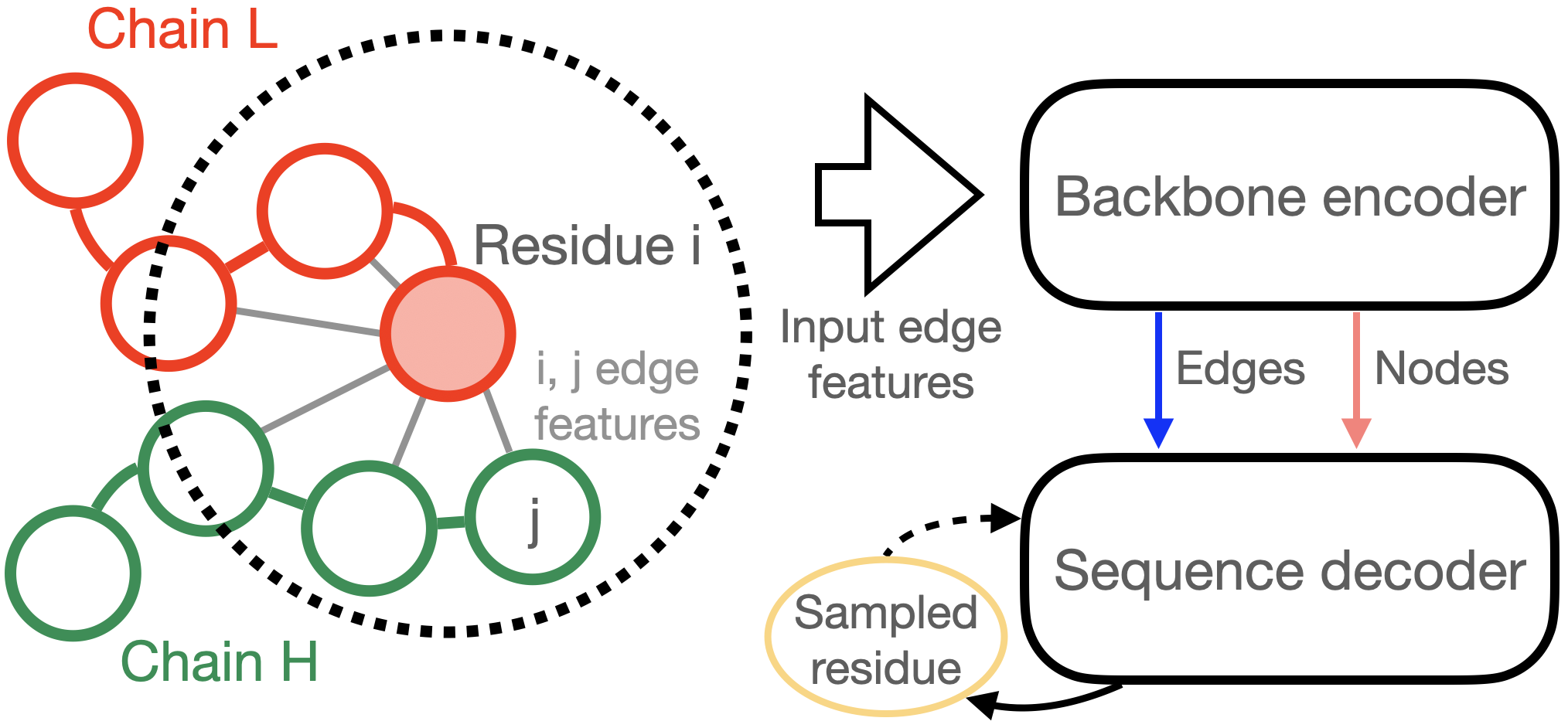}
    \caption{Schematic representation of the data processing steps and model architecture.}
    \label{fig:schema}
\end{figure}

We follow the recent approach introduced in the ProteinMPNN paper~\cite{proteinmpnn}, illustrated in Figure~\ref{fig:schema}.
This model is based on structured transformers using a message passing neural network (MPNN) as the aggregation function~\cite{structransfo} with the addition of an order agnostic decoding and edge updates in the encoder.
It aims to provide an autoregressive decomposition of the distribution of a protein sequence $\mathbf{s}$ given a backbone 3D structure $\mathbf{x}$
\begin{equation}
p(\mathbf{s}|\mathbf{x}) = \prod_i p(s_i| \mathbf{x}, s_{< i})
\end{equation}
where $p(s_i|\mathbf{x}, s_{< i})$ is the conditional probability of the amino acid $s_i$ at decoding step $i$, and $s_{<i} = \{s_1,\dots,s_{i-1}\}$ refers to previously decoded residues.\footnote{Note that this is not the same as their index in the chain, as residues are decoded in random order.}
These probabilities are parametrized using two components, an encoder that computes node and edge embeddings from structural information and a decoder that autoregressively predicts the next decoded residue given the preceding decoded letters and structural embeddings.

The backbone encoder takes as input distances between atoms as edge features, along with an all zero node vector.
The node features are updated by the three-layer MPNN. In contrast to~\citet{structransfo}, ProteinMPNN also then updates the edge features, before iterating through three layers of encoders.

The input edge features consist of the distances between $N$, $C_\alpha$, $C$, $O$, and virtual $C_\beta$ atoms of the 48 nearest residues in Euclidean space, decomposed in Gaussian radial basis functions. These inter-atom distance features are accompanied by a relative positional encoding in terms of the one-hot encoded distance in primary sequence space of two residues, with an additional token signaling whether they are in different chains.

A key change of ProteinMPNN to the original structured transformer implementation is the use of an order agnostic decoding, i.e.\ at each step the next residue to be predicted is chosen randomly among the remaining ones, with the full context of previous predictions given on either side.
Of particular relevance for antibodies with defined framework regions, this allows effective inference on structures with fixed regions where part of the sequence is known, which can then be provided as context.
The model is then trained to minimize the categorical cross entropy loss per residue.

\begin{figure*}
    \centering
    \includegraphics[width=0.555\linewidth]{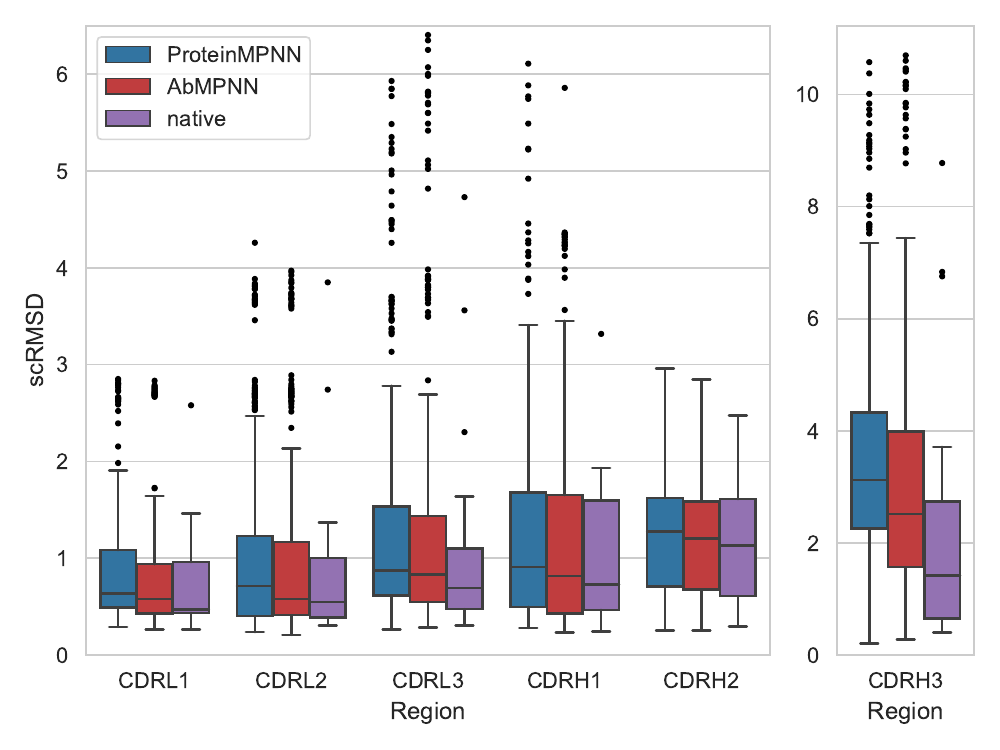}\qquad\;
    \includegraphics[width=0.281\linewidth]{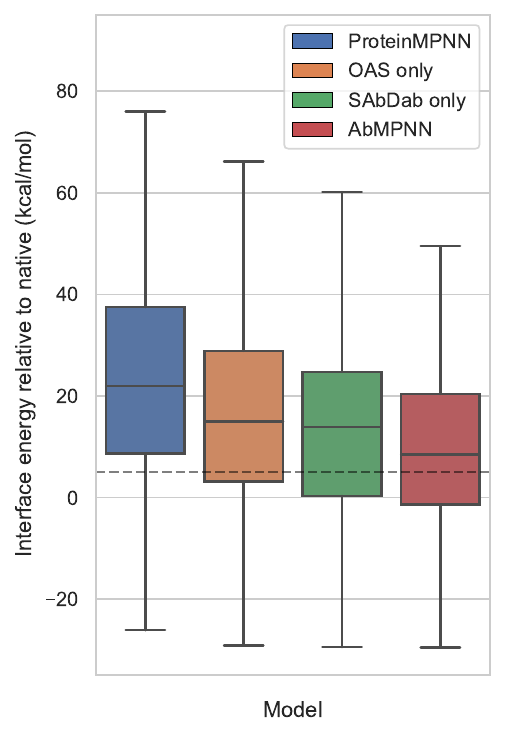}
    \caption{Left: Comparison of the backbone self-consistency RMSD between the original backbone and the ABodyBuilder2 structure predictions for the sequence obtained from ProteinMPNN (blue) and AbMPNN (red) as well as for the original sequence (purple).
    Right: Difference in interface energy as calculated by Rosetta of the heavy and light chains between each model prediction and the native sequence. The dashed line indicates a 5 kcal/mol threshold, and outliers are not displayed.}
    \label{fig:backbone}
\end{figure*}

\section{Training on antibody data}

Here we train an antibody specific variant of ProteinMPNN, which we will refer to as AbMPNN, that can predict valid antibody sequences and achieve improved accuracy in the variable region, notably for the CDR loops that determine antigen specificity.
We fine-tune the original ProteinMPNN model weights on antibody data. We consider two different datasets:
\begin{itemize}
    \item Antigen-binding fragments in complex from SAbDab. 
    We filter the full database for antibodies in complex with a protein antigen, and obtain 
    3500 complexes
    after removing redundant fragments and filtering out those with an experimental resolution worse than 5~\AA.
    \item 147919 paired heavy and light chain variable regions with unique concatenated CDRs from the OAS database. These are not experimentally resolved structures and do not contain epitopes, as the OAS contains only sequence data, but have been predicted using ABodyBuilder2 and structures are available as part of the ImmuneBuilder dataset~\cite{immunebuilder_data}.
\end{itemize}

We use the North definition of CDRs throughout this study~\cite{north}.
We fine-tune our model in two steps: first through an initial fine-tuning on the OAS structure predictions, for which we have a large number of natively paired heavy and light variable sequences modelled by ABodyBuilder2. We then further fine-tune the model on a small number of experimentally resolved antigen binding fragments in complex. In both cases the model is trained to predict the full variable domain, with the epitope given as context in the SAbDab training.


To filter the OAS antibodies, we first remove any duplicate entries with identical concatenated CDR sequences. For filtering the SAbDab antibody complexes, we remove antibodies that have both an identical concatenated CDR sequence and epitope sequences with greater than 90\% similarity by CD-HIT~\cite{cdhit}. Here the epitope is defined as the residues in each antigen chain with backbone atoms within 6~Å of the antibody backbone.

We next cluster the antibodies in both datasets using CD-HIT at 90\% similarity for the concatenated CDR sequences. This results in 107961 clusters for the OAS dataset, and 
1701 clusters for the SAbDab in complex dataset. Finally, we split these into training, validation and test sets, with a ratio of 8-1-1. To ensure that there is no pollution between the OAS and SAbDab dataset, we ensure that any cluster with a sequence in the SAbDab dataset that would have been clustered into a OAS training or validation cluster is placed into the SAbDab training set\footnote{This corresponds to 50 SAbDab clusters before filtering for resolution.}.

We fine-tune our AbMPNN model starting from the original ProteinMPNN model weights, first on the OAS dataset, then on the SAbDab dataset. Both of these fine-tuning steps use an Adam optimiser. We reduce the learning rate by a factor of 10 if the validation loss does not improve for 10 epochs (5 for the OAS data), starting from an initial learning rate of $5 \times 10^{-4}$ for the OAS fine-tuning and of $10^{-4}$ for the SAbDab training step. Each epoch consists of 1000 randomly selected antigen binding fragments.

\begin{figure*}
    \centering
    \includegraphics[width=0.5\linewidth]{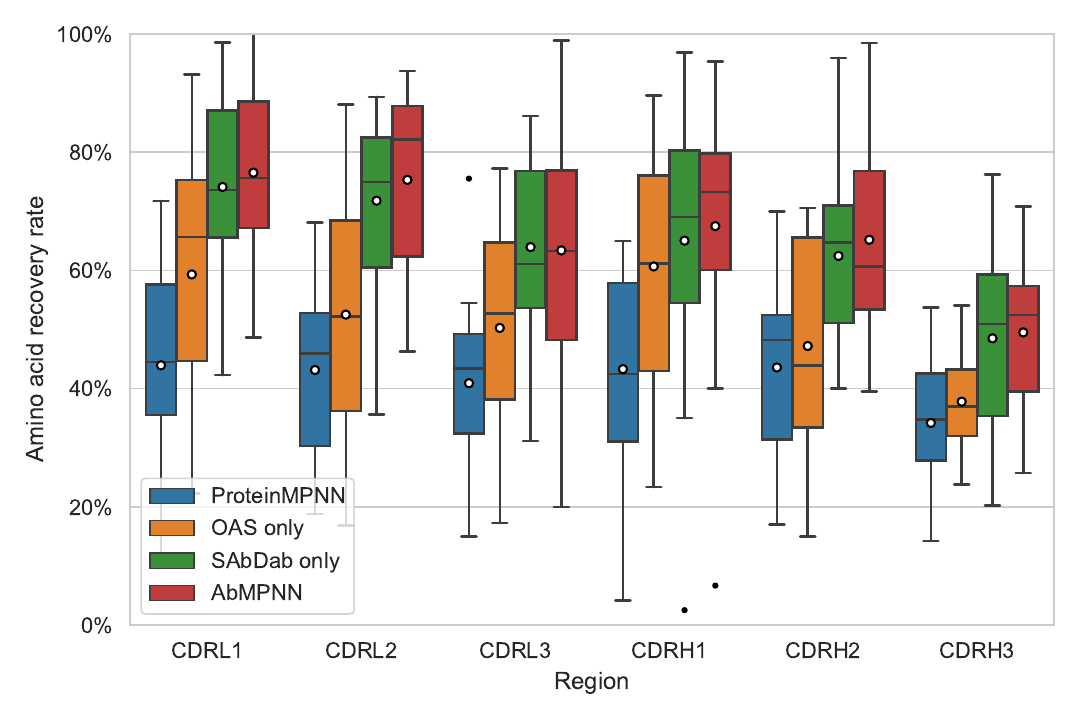}%
    \includegraphics[width=0.5\linewidth]{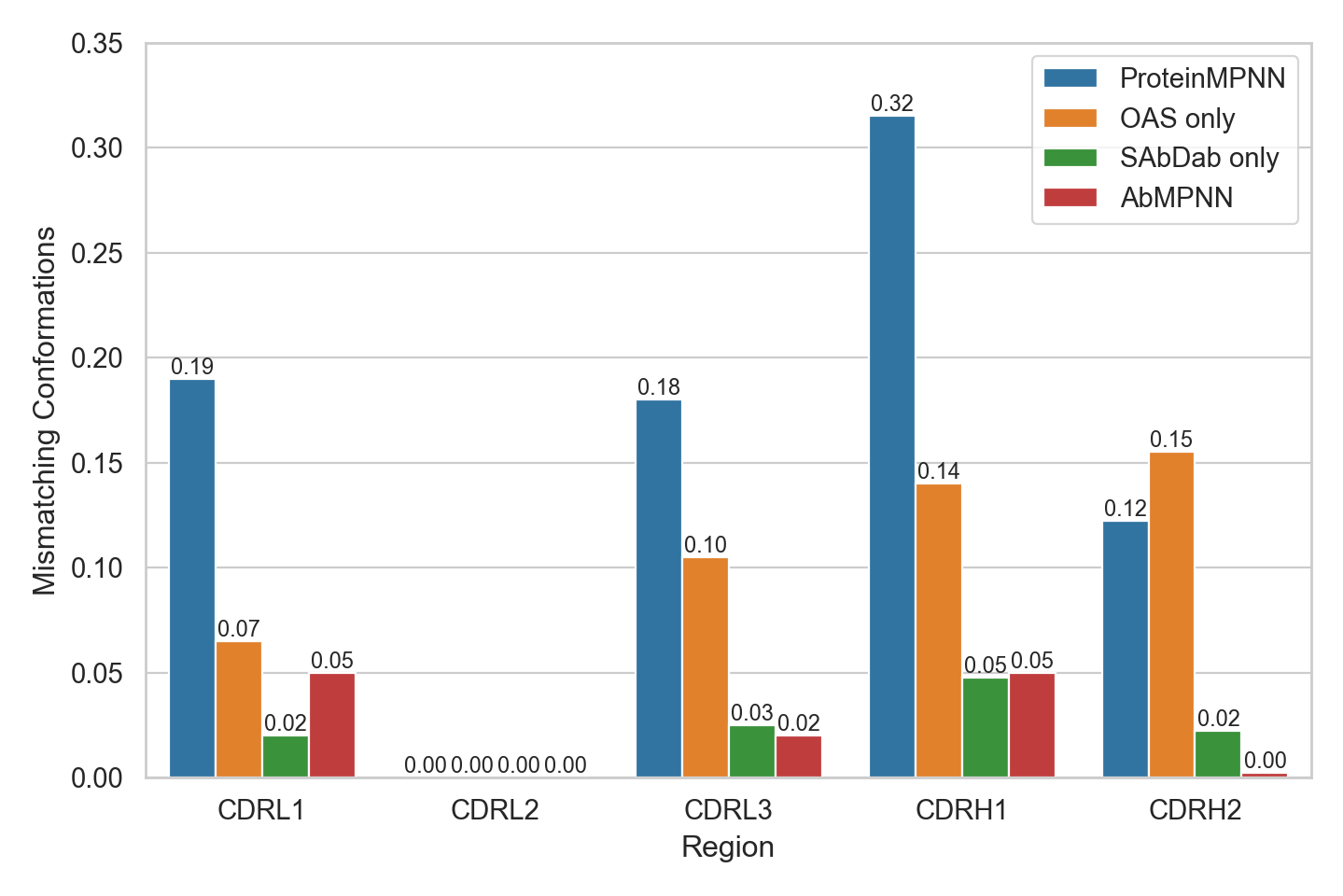}
    \caption{Left: Sequence recovery of the original ProteinMPNN model and of  AbMPNN, shown separately for each CDR loop with the mean value indicated by a white circle. Right: Fraction of mismatching conformations as predicted by SCALOP for CDR L1-3 and CDR H1-2.}
    \label{fig:benchmarks}
\end{figure*}

\section{Designability and accuracy study}

We now report on the results obtained from our fine-tuned AbMPNN model, and analyze its accuracy and robustness.
For each model, we run inference to predict 20 variations of the CDR sequences for 20 complexes from different clusters in the SAbDab testset, using a sampling temperature of 0.2 at inference.

We start by studying the designability of sequences predicted by AbMPNN. To this end, we assess the self-consistency of the model similarly to \citet{trippe2022diffusion}, by giving the sequences as input to a structure prediction model, in this case ABodyBuilder2, and compute the RMSD between the original backbone and the predicted structure.
This is shown in Figure~\ref{fig:backbone} (left), where the RMSD is shown separately for each CDR loop. Here the self-consistency RMSD for the original sequence is displayed in purple, and provides a benchmark of how well a model can perform on this designability test. The boxes show the first and third quartile, with the median shown as a horizontal line, and outliers\footnote{Outliers are defined as values more than 1.5 times the interquartile range away from the first or third quartile.} indicated as dots. Both the original ProteinMPNN model and the fine-tuned AbMPNN are shown, and we can observe a clear improvement of about ~20\% on the median RMSD of CDR-H3 for the fine-tuned model, with smaller improvements visible for all other CDRs.

To further probe designability, we compute the interface energy of the heavy and light chains of these structures using Rosetta~\cite{rosetta}. The difference of the interface energy for each model with the structure predicted from the native sequence is shown in Figure~\ref{fig:backbone} (right), where we also display intermediate models fine-tuned on only one of the antibody datasets.\footnote{For readability reasons, we do not display the $\sim 5\%$ percent of outliers with sometimes very large interface energy values.} Here we note that the antibody interface energy improves after each fine-tuning step, with our final AbMPNN model producing structures predicted to be more stable than the original ProteinMPNN model. Interestingly, 40\% of AbMPNN sequences are within 5 kcal/mol or less of the original sequence interface energy, compared with only 20.5\% for ProteinMPNN, indicating that the AbMPNN model is better at recovering light and heavy chain residue contacts and generating stable heavy chain - light chain dimers.

In Figure~\ref{fig:benchmarks} (left), we show the proportion of correctly recovered residues across each CDR loop. A notable improvement can again be seen for the fine-tuned model, with sequence recovery rates around 60\% for the AbMPNN model while the original ProteinMPNN model only has a recovery rate of about 40\% across CDR loops.
Despite having a relatively low RMSD with the input structure when predicted using ABodyBuilder2, the sequences predicted by the original ProteinMPNN model have significant differences to the original antibody sequences. 

We examine the conformations of the CDR loops. We cluster the non-H3 CDRs into canonical forms using SCALOP~\cite{scalop}. Canonical clusters are built from sequence space using position-specific scoring matrices. The fraction of predicted sequences that do not match the canonical form of their original sequence is shown in Figure~\ref{fig:benchmarks} (right). Here we observe a large improvement in recovery of the correct canonical cluster with the fine-tuned AbMPNN model.

We consider the validity of the predicted antibody sequences, by redesigning the full variable region of our previous test set, this time including the framework region.
We then annotate these sequences using ANARCI~\cite{anarci}, a tool for numbering variable domains, and find that while every sequence predicted by AbMPNN is recognised as an antibody sequence, 16.8\% of those predicted by ProteinMPNN can not be annotated due to errors in the framework region of either the light or heavy chain.

\section{Conclusions}

In this article, we have introduced an inverse folding model specifically adapted to antibodies to predict their sequences based on structural backbone information. This model follows the architecture of the recent generic protein model ProteinMPNN, and is fine-tuned on experimental structures of antigen binding fragments as well as numerical structure predictions of variable antibody fragments derived from the Observed Antibody Space.
We showed that with a few changes and appropriate retraining, our AbMPNN model can set new state-of-the-art benchmarks for designability and amino acid sequence recovery, particularly for the hypervariable CDR-H3 loop.
We discussed the canonical forms of CDRs and showed that a sequence-based conformational clustering achieves excellent recovery of the original sequence cluster for all available CDR loops.
Antibody-specific inverse folding tools can provide a powerful approach to AI-driven drug discovery, notably by improving designability and affinity of existing binders, and as a final sequence recovery step for de novo structural models~\cite{rfdiffusion}.
We release the weights of our model to allow for the use of this work in other downstream applications~\cite{frederic_a_dreyer_2023_8164693}.



\section*{Acknowledgements}
We are grateful to Aleksandr Kovaltsuk, Newton Wahome and Jody Barbeau for helpful suggestions and detailed comments on the manuscript, and to Claire Marks for useful discussions.

\bibliography{references}

\begin{thebibliography}{28}
\providecommand{\natexlab}[1]{#1}
\providecommand{\url}[1]{\texttt{#1}}
\expandafter\ifx\csname urlstyle\endcsname\relax
  \providecommand{\doi}[1]{doi: #1}\else
  \providecommand{\doi}{doi: \begingroup \urlstyle{rm}\Url}\fi

\bibitem[Abanades(2022)]{immunebuilder_data}
Abanades, B.
\newblock {ImmuneBuilder: Deep-Learning models for predicting the structures of
  immune proteins.}, November 2022.
\newblock URL \url{https://doi.org/10.5281/zenodo.7258553}.

\bibitem[Abanades et~al.(2022)Abanades, Wong, Boyles, Georges, Bujotzek, and
  Deane]{abb2}
Abanades, B., Wong, W.~K., Boyles, F., Georges, G., Bujotzek, A., and Deane,
  C.~M.
\newblock Immunebuilder: Deep-learning models for predicting the structures of
  immune proteins.
\newblock \emph{bioRxiv}, 2022.
\newblock \doi{10.1101/2022.11.04.514231}.
\newblock URL
  \url{https://www.biorxiv.org/content/early/2022/11/04/2022.11.04.514231}.

\bibitem[Alford et~al.(2017)Alford, Leaver-Fay, Jeliazkov, O'Meara, DiMaio,
  Park, Shapovalov, Renfrew, Mulligan, Kappel, Labonte, Pacella, Bonneau,
  Bradley, Dunbrack, Das, Baker, Kuhlman, Kortemme, and Gray]{rosetta}
Alford, R., Leaver-Fay, A., Jeliazkov, J., O'Meara, M., DiMaio, F., Park, H.,
  Shapovalov, M., Renfrew, P., Mulligan, V., Kappel, K., Labonte, J., Pacella,
  M., Bonneau, R., Bradley, P., Dunbrack, R., Das, R., Baker, D., Kuhlman, B.,
  Kortemme, T., and Gray, J.
\newblock The rosetta all-atom energy function for macromolecular modeling and
  design.
\newblock \emph{Journal of Chemical Theory and Computation}, 13\penalty0
  (6):\penalty0 3031--3048, June 2017.
\newblock ISSN 1549-9618.
\newblock \doi{10.1021/acs.jctc.7b00125}.

\bibitem[Anand-Achim et~al.(2021)Anand-Achim, Eguchi, Mathews, Perez, Derry,
  Altman, and Huang]{AnandAchim}
Anand-Achim, N., Eguchi, R.~R., Mathews, I.~I., Perez, C.~P., Derry, A.,
  Altman, R.~B., and Huang, P.-S.
\newblock Protein sequence design with a learned potential.
\newblock \emph{bioRxiv}, 2021.
\newblock \doi{10.1101/2020.01.06.895466}.
\newblock URL
  \url{https://www.biorxiv.org/content/early/2021/03/02/2020.01.06.895466}.

\bibitem[Baek et~al.(2021)Baek, DiMaio, Anishchenko, Dauparas, Ovchinnikov,
  Lee, Wang, Cong, Kinch, Schaeffer, Millán, Park, Adams, Glassman,
  DeGiovanni, Pereira, Rodrigues, van Dijk, Ebrecht, Opperman, Sagmeister,
  Buhlheller, Pavkov-Keller, Rathinaswamy, Dalwadi, Yip, Burke, Garcia,
  Grishin, Adams, Read, and Baker]{rosettafold}
Baek, M., DiMaio, F., Anishchenko, I., Dauparas, J., Ovchinnikov, S., Lee,
  G.~R., Wang, J., Cong, Q., Kinch, L.~N., Schaeffer, R.~D., Millán, C., Park,
  H., Adams, C., Glassman, C.~R., DeGiovanni, A., Pereira, J.~H., Rodrigues,
  A.~V., van Dijk, A.~A., Ebrecht, A.~C., Opperman, D.~J., Sagmeister, T.,
  Buhlheller, C., Pavkov-Keller, T., Rathinaswamy, M.~K., Dalwadi, U., Yip,
  C.~K., Burke, J.~E., Garcia, K.~C., Grishin, N.~V., Adams, P.~D., Read,
  R.~J., and Baker, D.
\newblock Accurate prediction of protein structures and interactions using a
  three-track neural network.
\newblock \emph{Science}, 373\penalty0 (6557):\penalty0 871--876, 2021.
\newblock \doi{10.1126/science.abj8754}.
\newblock URL \url{https://www.science.org/doi/abs/10.1126/science.abj8754}.

\bibitem[Dauparas et~al.(2022)Dauparas, Anishchenko, Bennett, Bai, Ragotte,
  Milles, Wicky, Courbet, de~Haas, Bethel, Leung, Huddy, Pellock, Tischer,
  Chan, Koepnick, Nguyen, Kang, Sankaran, Bera, King, and Baker]{proteinmpnn}
Dauparas, J., Anishchenko, I., Bennett, N., Bai, H., Ragotte, R.~J., Milles,
  L.~F., Wicky, B. I.~M., Courbet, A., de~Haas, R.~J., Bethel, N., Leung, P.
  J.~Y., Huddy, T.~F., Pellock, S., Tischer, D., Chan, F., Koepnick, B.,
  Nguyen, H., Kang, A., Sankaran, B., Bera, A.~K., King, N.~P., and Baker, D.
\newblock Robust deep learning based protein sequence design using proteinmpnn.
\newblock \emph{bioRxiv}, 2022.
\newblock \doi{10.1101/2022.06.03.494563}.
\newblock URL
  \url{https://www.biorxiv.org/content/early/2022/06/04/2022.06.03.494563}.

\bibitem[Dreyer et~al.(2023)Dreyer, Cutting, Schneider, Kenlay, and
  Deane]{frederic_a_dreyer_2023_8164693}
Dreyer, F.~A., Cutting, D., Schneider, C., Kenlay, H., and Deane, C.~M.
\newblock {Inverse folding for antibody sequence design using deep learning},
  July 2023.
\newblock URL \url{https://doi.org/10.5281/zenodo.8164693}.

\bibitem[Dunbar \& Deane(2015)Dunbar and Deane]{anarci}
Dunbar, J. and Deane, C.~M.
\newblock {ANARCI: antigen receptor numbering and receptor classification}.
\newblock \emph{Bioinformatics}, 32\penalty0 (2):\penalty0 298--300, 09 2015.
\newblock ISSN 1367-4803.
\newblock \doi{10.1093/bioinformatics/btv552}.
\newblock URL \url{https://doi.org/10.1093/bioinformatics/btv552}.

\bibitem[Dunbar et~al.(2013)Dunbar, Krawczyk, Leem, Baker, Fuchs, Georges, Shi,
  and Deane]{sabdab1}
Dunbar, J., Krawczyk, K., Leem, J., Baker, T., Fuchs, A., Georges, G., Shi, J.,
  and Deane, C.~M.
\newblock {SAbDab: the structural antibody database}.
\newblock \emph{Nucleic Acids Research}, 42\penalty0 (D1):\penalty0
  D1140--D1146, 11 2013.
\newblock ISSN 0305-1048.
\newblock \doi{10.1093/nar/gkt1043}.
\newblock URL \url{https://doi.org/10.1093/nar/gkt1043}.

\bibitem[Fu et~al.(2012)Fu, Niu, Zhu, Wu, and Li]{cdhit}
Fu, L., Niu, B., Zhu, Z., Wu, S., and Li, W.
\newblock {CD-HIT: accelerated for clustering the next-generation sequencing
  data}.
\newblock \emph{Bioinformatics}, 28\penalty0 (23):\penalty0 3150--3152, 10
  2012.
\newblock ISSN 1367-4803.
\newblock \doi{10.1093/bioinformatics/bts565}.
\newblock URL \url{https://doi.org/10.1093/bioinformatics/bts565}.

\bibitem[Hsu et~al.(2022)Hsu, Verkuil, Liu, Lin, Hie, Sercu, Lerer, and
  Rives]{esmif}
Hsu, C., Verkuil, R., Liu, J., Lin, Z., Hie, B., Sercu, T., Lerer, A., and
  Rives, A.
\newblock Learning inverse folding from millions of predicted structures.
\newblock \emph{bioRxiv}, 2022.
\newblock \doi{10.1101/2022.04.10.487779}.
\newblock URL
  \url{https://www.biorxiv.org/content/early/2022/04/10/2022.04.10.487779}.

\bibitem[Ingraham et~al.(2019)Ingraham, Garg, Barzilay, and
  Jaakkola]{structransfo}
Ingraham, J., Garg, V., Barzilay, R., and Jaakkola, T.
\newblock Generative models for graph-based protein design.
\newblock In Wallach, H., Larochelle, H., Beygelzimer, A., d\textquotesingle
  Alch\'{e}-Buc, F., Fox, E., and Garnett, R. (eds.), \emph{Advances in Neural
  Information Processing Systems}, volume~32. Curran Associates, Inc., 2019.
\newblock URL
  \url{https://proceedings.neurips.cc/paper_files/paper/2019/file/f3a4ff4839c56a5f460c88cce3666a2b-Paper.pdf}.

\bibitem[Jing et~al.(2021)Jing, Eismann, Suriana, Townshend, and
  Dror]{jing2021learning}
Jing, B., Eismann, S., Suriana, P., Townshend, R. J.~L., and Dror, R.
\newblock Learning from protein structure with geometric vector perceptrons.
\newblock In \emph{International Conference on Learning Representations}, 2021.
\newblock URL \url{https://openreview.net/forum?id=1YLJDvSx6J4}.

\bibitem[Jumper et~al.(2021)Jumper, Evans, Pritzel, Green, Figurnov,
  Ronneberger, Tunyasuvunakool, Bates, {\v Z}{\'\i}dek, Potapenko, Bridgland,
  Meyer, Kohl, Ballard, Cowie, Romera-Paredes, Nikolov, Jain, Adler, Back,
  Petersen, Reiman, Clancy, Zielinski, Steinegger, Pacholska, Berghammer,
  Bodenstein, Silver, Vinyals, Senior, Kavukcuoglu, Kohli, and Hassabis]{af2}
Jumper, J., Evans, R., Pritzel, A., Green, T., Figurnov, M., Ronneberger, O.,
  Tunyasuvunakool, K., Bates, R., {\v Z}{\'\i}dek, A., Potapenko, A.,
  Bridgland, A., Meyer, C., Kohl, S. A.~A., Ballard, A.~J., Cowie, A.,
  Romera-Paredes, B., Nikolov, S., Jain, R., Adler, J., Back, T., Petersen, S.,
  Reiman, D., Clancy, E., Zielinski, M., Steinegger, M., Pacholska, M.,
  Berghammer, T., Bodenstein, S., Silver, D., Vinyals, O., Senior, A.~W.,
  Kavukcuoglu, K., Kohli, P., and Hassabis, D.
\newblock Highly accurate protein structure prediction with alphafold.
\newblock \emph{Nature}, 596\penalty0 (7873):\penalty0 583--589, 2021.
\newblock \doi{10.1038/s41586-021-03819-2}.
\newblock URL \url{https://doi.org/10.1038/s41586-021-03819-2}.

\bibitem[Kovaltsuk et~al.(2018)Kovaltsuk, Leem, Kelm, Snowden, Deane, and
  Krawczyk]{oas1}
Kovaltsuk, A., Leem, J., Kelm, S., Snowden, J., Deane, C.~M., and Krawczyk, K.
\newblock {Observed Antibody Space: A Resource for Data Mining Next-Generation
  Sequencing of Antibody Repertoires}.
\newblock \emph{The Journal of Immunology}, 201\penalty0 (8):\penalty0
  2502--2509, 10 2018.
\newblock ISSN 0022-1767.
\newblock \doi{10.4049/jimmunol.1800708}.
\newblock URL \url{https://doi.org/10.4049/jimmunol.1800708}.

\bibitem[Lin \& AlQuraishi(2023)Lin and AlQuraishi]{genie}
Lin, Y. and AlQuraishi, M.
\newblock Generating novel, designable, and diverse protein structures by
  equivariantly diffusing oriented residue clouds, 2023.

\bibitem[Lin et~al.(2022)Lin, Akin, Rao, Hie, Zhu, Lu, dos Santos~Costa,
  Fazel-Zarandi, Sercu, Candido, and Rives]{esmfold}
Lin, Z., Akin, H., Rao, R., Hie, B., Zhu, Z., Lu, W., dos Santos~Costa, A.,
  Fazel-Zarandi, M., Sercu, T., Candido, S., and Rives, A.
\newblock Language models of protein sequences at the scale of evolution enable
  accurate structure prediction.
\newblock \emph{bioRxiv}, 2022.
\newblock \doi{10.1101/2022.07.20.500902}.
\newblock URL
  \url{https://www.biorxiv.org/content/early/2022/07/21/2022.07.20.500902}.

\bibitem[Narciso et~al.(2011)Narciso, Uy, Cabang, Chavez, Pablo,
  Padilla-Concepcion, and Padlan]{cdrh3_2011}
Narciso, J.~E., Uy, I., Cabang, A., Chavez, J., Pablo, J., Padilla-Concepcion,
  G., and Padlan, E.
\newblock Analysis of the antibody structure based on high-resolution
  crystallographic studies.
\newblock \emph{New biotechnology}, 28:\penalty0 435--47, 04 2011.
\newblock \doi{10.1016/j.nbt.2011.03.012}.

\bibitem[North et~al.(2011)North, Lehmann, and Dunbrack]{north}
North, B., Lehmann, A., and Dunbrack, R.~L.
\newblock A new clustering of antibody cdr loop conformations.
\newblock \emph{Journal of Molecular Biology}, 406\penalty0 (2):\penalty0
  228--256, 2011.
\newblock ISSN 0022-2836.
\newblock \doi{https://doi.org/10.1016/j.jmb.2010.10.030}.
\newblock URL
  \url{https://www.sciencedirect.com/science/article/pii/S0022283610011496}.

\bibitem[Olsen et~al.(2022)Olsen, Boyles, and Deane]{oas2}
Olsen, T.~H., Boyles, F., and Deane, C.~M.
\newblock Observed antibody space: A diverse database of cleaned, annotated,
  and translated unpaired and paired antibody sequences.
\newblock \emph{Protein Science}, 31\penalty0 (1):\penalty0 141--146, 2022.
\newblock \doi{https://doi.org/10.1002/pro.4205}.
\newblock URL \url{https://onlinelibrary.wiley.com/doi/abs/10.1002/pro.4205}.

\bibitem[Schneider et~al.(2022)Schneider, Raybould, and Deane]{sabdab2}
Schneider, C., Raybould, M. I.~J., and Deane, C.~M.
\newblock {SAbDab in the age of biotherapeutics: updates including SAbDab-nano,
  the nanobody structure tracker}.
\newblock \emph{Nucleic Acids Research}, 50\penalty0 (D1):\penalty0
  D1368--D1372, 02 2022.
\newblock ISSN 0305-1048.
\newblock \doi{10.1093/nar/gkab1050}.
\newblock URL \url{https://doi.org/10.1093/nar/gkab1050}.

\bibitem[Sela-Culang et~al.(2013)Sela-Culang, Kunik, and Ofran]{cdrs}
Sela-Culang, I., Kunik, V., and Ofran, Y.
\newblock The structural basis of antibody-antigen recognition.
\newblock \emph{Frontiers in Immunology}, 4\penalty0 (OCT), October 2013.
\newblock ISSN 1664-3224.
\newblock \doi{https://doi.org/10.3389/fimmu.2013.00302}.

\bibitem[Strokach et~al.(2020)Strokach, Becerra, Corbi-Verge, Perez-Riba, and
  Kim]{STROKACH2020402}
Strokach, A., Becerra, D., Corbi-Verge, C., Perez-Riba, A., and Kim, P.~M.
\newblock Fast and flexible protein design using deep graph neural networks.
\newblock \emph{Cell Systems}, 11\penalty0 (4):\penalty0 402--411.e4, 2020.
\newblock ISSN 2405-4712.
\newblock \doi{https://doi.org/10.1016/j.cels.2020.08.016}.
\newblock URL
  \url{https://www.sciencedirect.com/science/article/pii/S2405471220303276}.

\bibitem[Trippe et~al.(2022)Trippe, Yim, Tischer, Baker, Broderick, Barzilay,
  and Jaakkola]{trippe2022diffusion}
Trippe, B.~L., Yim, J., Tischer, D., Baker, D., Broderick, T., Barzilay, R.,
  and Jaakkola, T.
\newblock Diffusion probabilistic modeling of protein backbones in 3d for the
  motif-scaffolding problem, 2022.

\bibitem[Tsuchiya \& Mizuguchi(2016)Tsuchiya and Mizuguchi]{cdrh3}
Tsuchiya, Y. and Mizuguchi, K.
\newblock The diversity of h3 loops determines the antigen-binding tendencies
  of antibody cdr loops.
\newblock \emph{Protein science : a publication of the Protein Society}, 25, 01
  2016.
\newblock \doi{10.1002/pro.2874}.

\bibitem[Watson et~al.(2023)Watson, Juergens, Bennett, Trippe, Yim, Eisenach,
  Ahern, Borst, Ragotte, Milles, Wicky, Hanikel, Pellock, Courbet, Sheffler,
  Wang, Venkatesh, Sappington, Torres, Lauko, De~Bortoli, Mathieu, Ovchinnikov,
  Barzilay, Jaakkola, DiMaio, Baek, and Baker]{rfdiffusion}
Watson, J.~L., Juergens, D., Bennett, N.~R., Trippe, B.~L., Yim, J., Eisenach,
  H.~E., Ahern, W., Borst, A.~J., Ragotte, R.~J., Milles, L.~F., Wicky, B.
  I.~M., Hanikel, N., Pellock, S.~J., Courbet, A., Sheffler, W., Wang, J.,
  Venkatesh, P., Sappington, I., Torres, S.~V., Lauko, A., De~Bortoli, V.,
  Mathieu, E., Ovchinnikov, S., Barzilay, R., Jaakkola, T.~S., DiMaio, F.,
  Baek, M., and Baker, D.
\newblock De novo design of protein structure and function with rfdiffusion.
\newblock \emph{Nature}, 2023.
\newblock \doi{10.1038/s41586-023-06415-8}.
\newblock URL \url{https://doi.org/10.1038/s41586-023-06415-8}.

\bibitem[Wong et~al.(2018)Wong, Georges, Ros, Kelm, Lewis, Taddese, Leem, and
  Deane]{scalop}
Wong, W.~K., Georges, G., Ros, F., Kelm, S., Lewis, A.~P., Taddese, B., Leem,
  J., and Deane, C.~M.
\newblock {SCALOP: sequence-based antibody canonical loop structure
  annotation}.
\newblock \emph{Bioinformatics}, 35\penalty0 (10):\penalty0 1774--1776, 10
  2018.
\newblock ISSN 1367-4803.
\newblock \doi{10.1093/bioinformatics/bty877}.
\newblock URL \url{https://doi.org/10.1093/bioinformatics/bty877}.

\bibitem[Yim et~al.(2023)Yim, Trippe, Bortoli, Mathieu, Doucet, Barzilay, and
  Jaakkola]{framediff}
Yim, J., Trippe, B.~L., Bortoli, V.~D., Mathieu, E., Doucet, A., Barzilay, R.,
  and Jaakkola, T.
\newblock Se(3) diffusion model with application to protein backbone
  generation, 2023.

\end{thebibliography}
\bibliographystyle{icml2023}

\newpage
\appendix
\onecolumn
\section{Study of germline distributions}
\begin{figure*}[ht]
    \centering
    \includegraphics[width=0.5\linewidth]{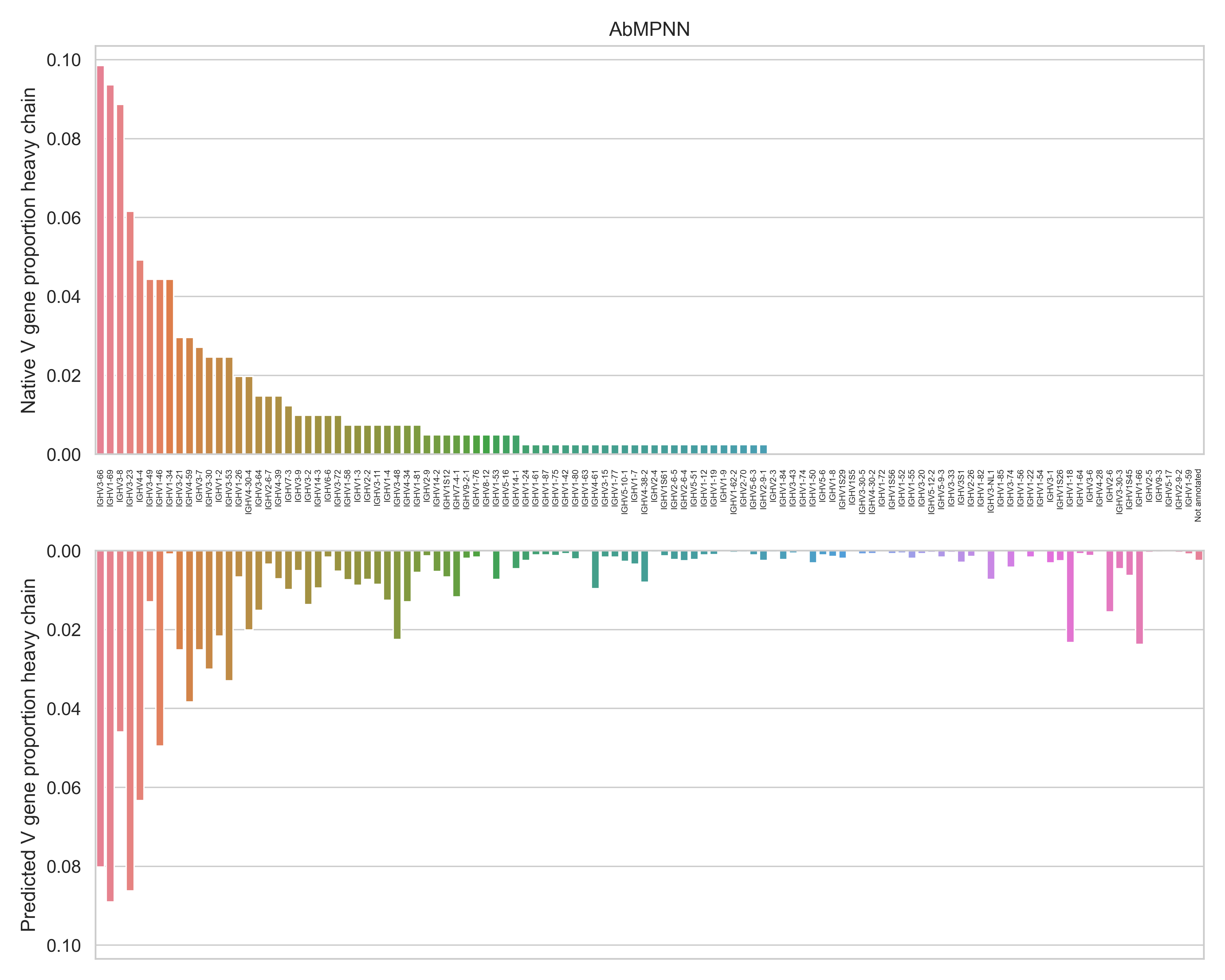}%
    \includegraphics[width=0.5\linewidth]{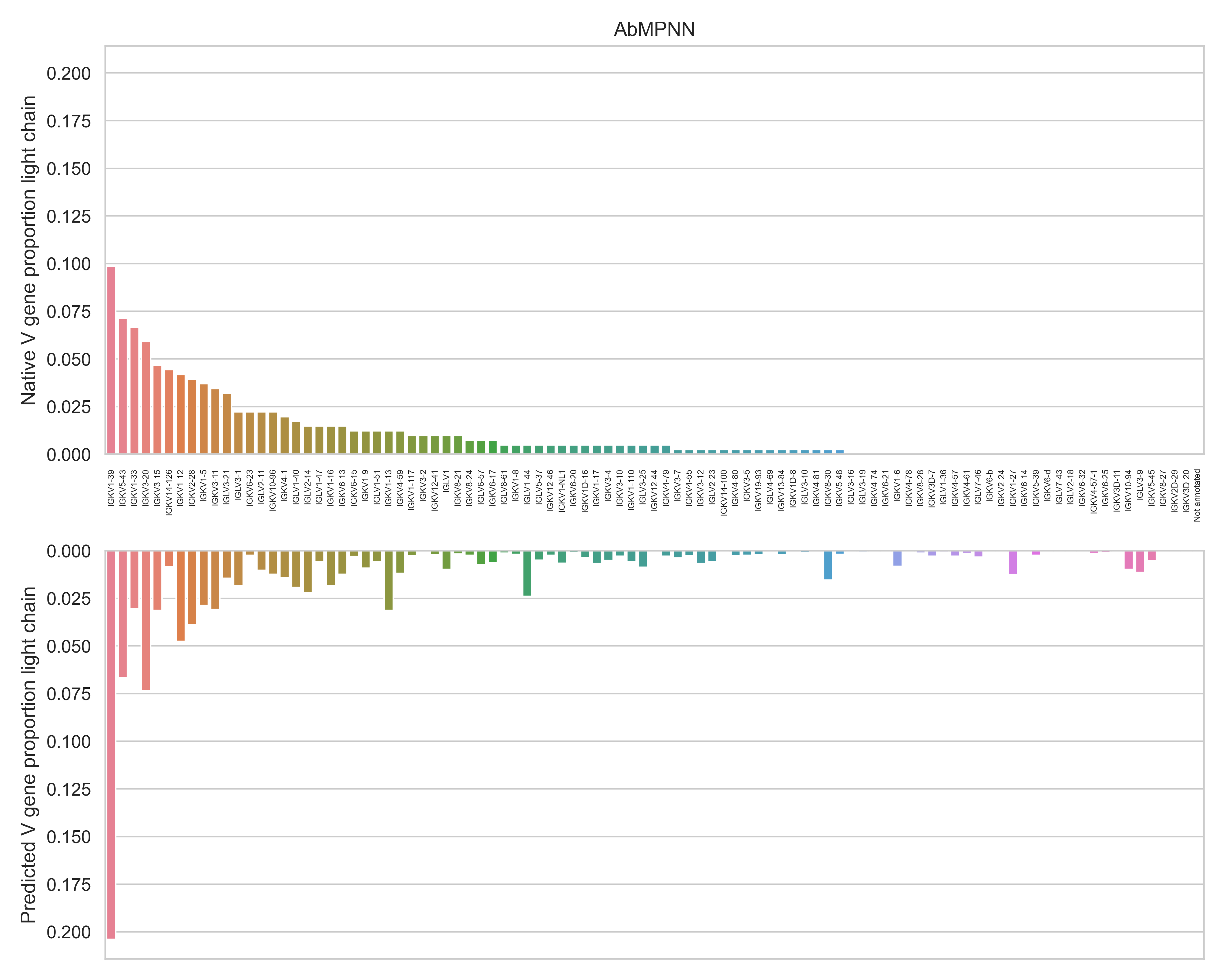}
    \caption{Comparison of the V-gene germline assignments from the SAbDab dataset and in corresponding predictions from AbMPNN for both the heavy (left) and light (right) chains.}
    \label{fig:vgenedist}
\end{figure*}

\begin{figure*}[ht]
    \centering
    \includegraphics[width=0.5\linewidth]{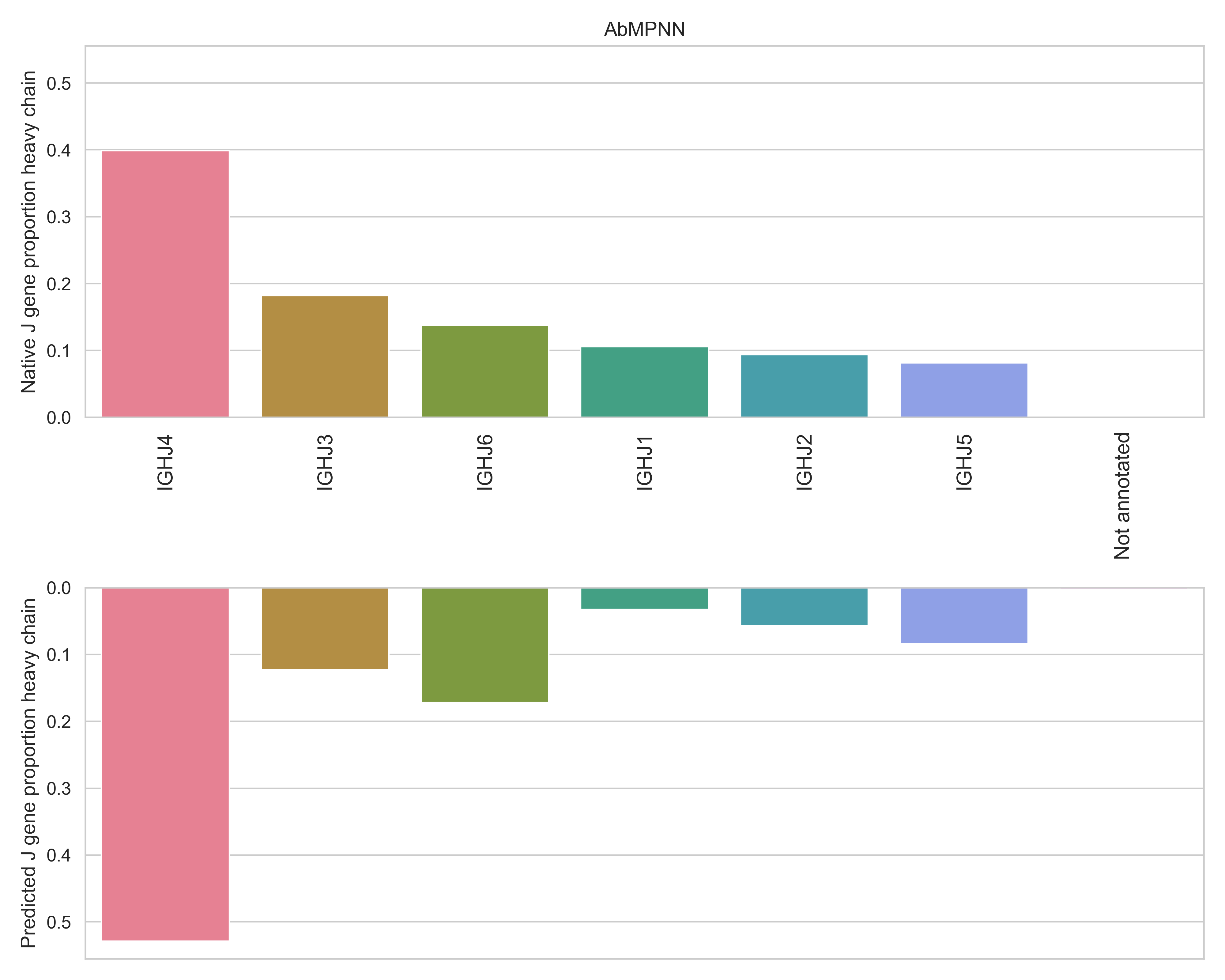}%
    \includegraphics[width=0.5\linewidth]{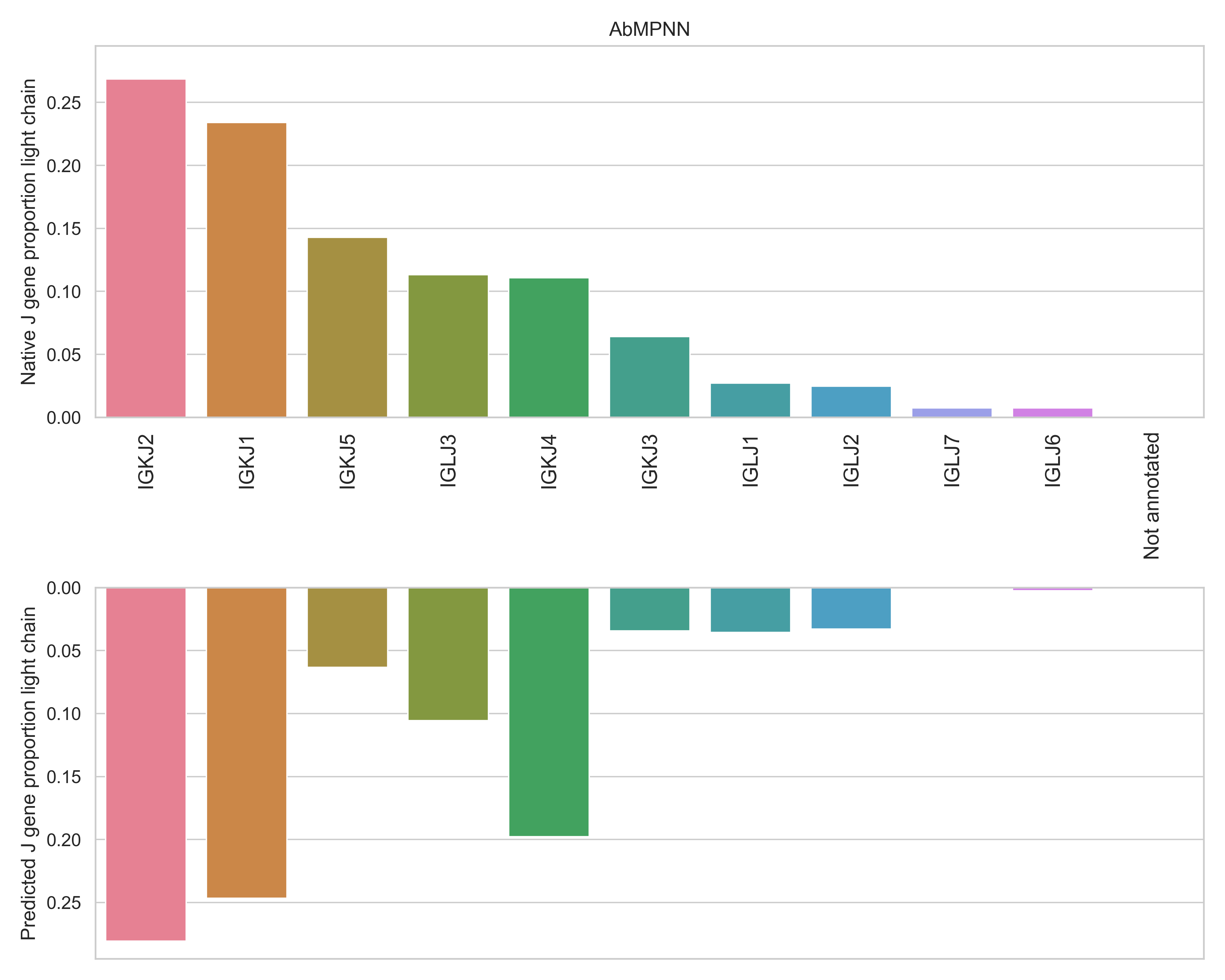}
    \caption{Comparison of the J-gene germline assignments from the SAbDab dataset and in corresponding predictions from AbMPNN for both the heavy (left) and light (right) chains.}
    \label{fig:jgenedist}
\end{figure*}

\begin{figure*}[ht]
    \centering
    \includegraphics[width=0.5\linewidth]{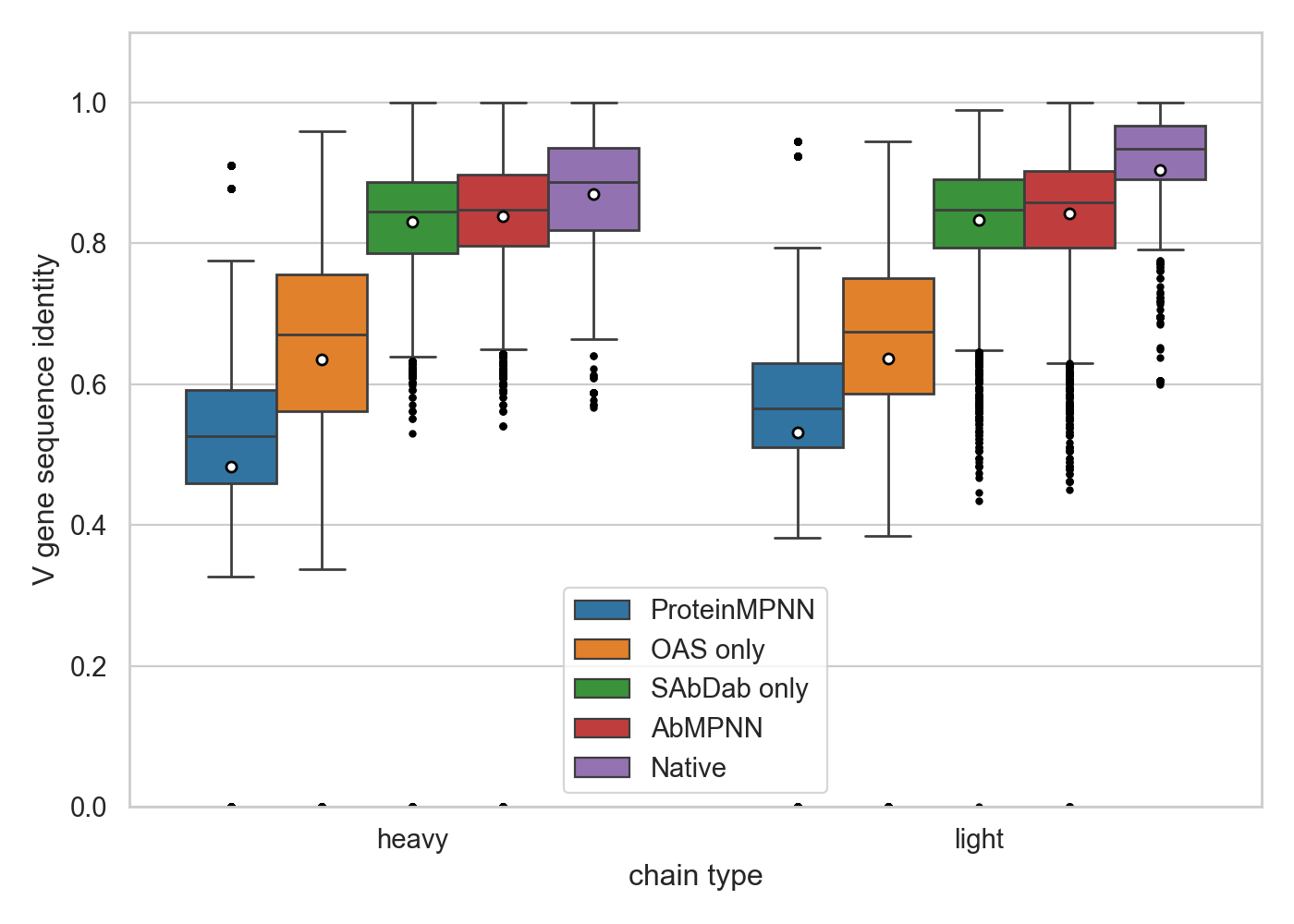}%
    \includegraphics[width=0.5\linewidth]{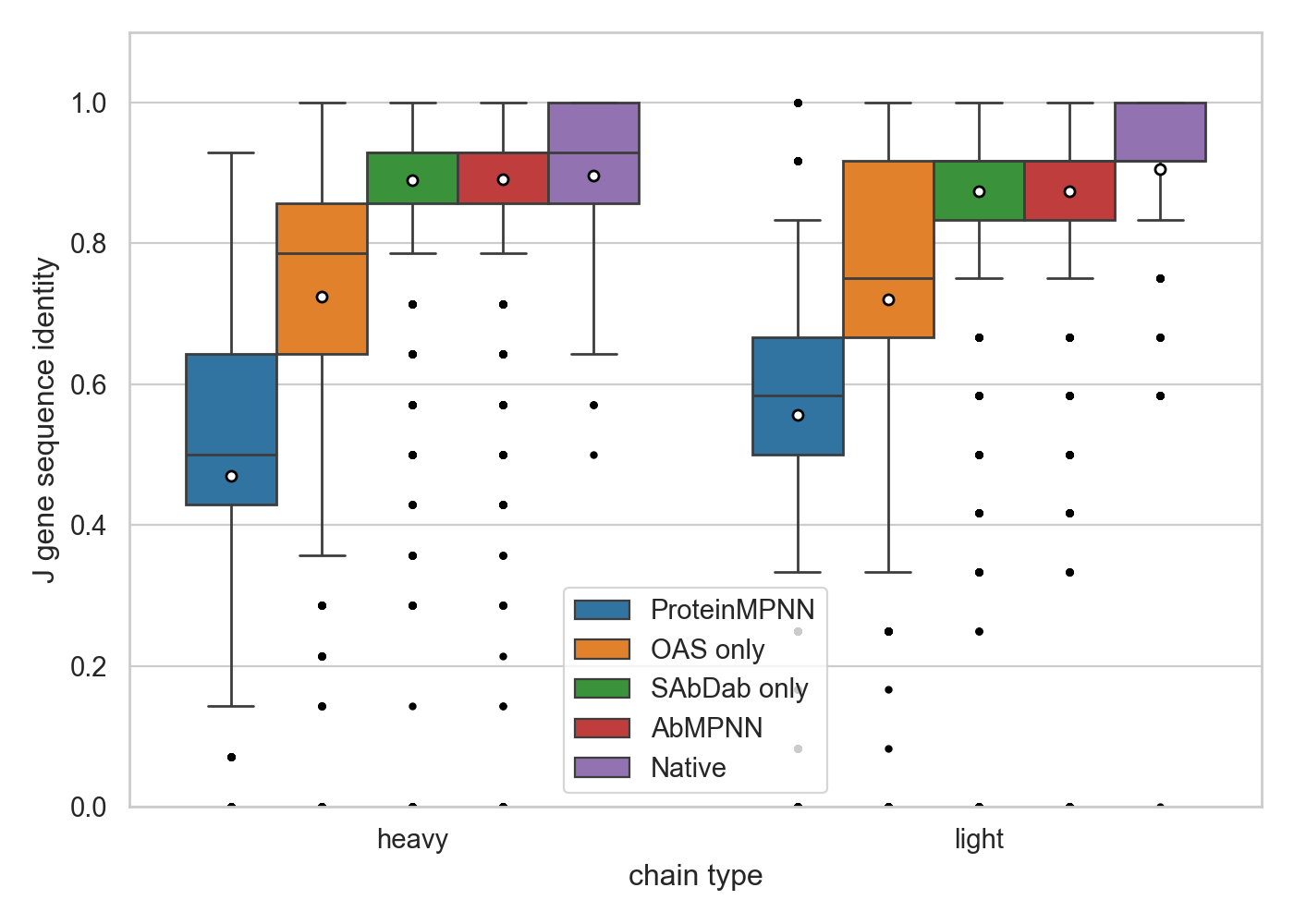}
    \caption{Sequence identity with the closest V-gene (left) or J-gene (right) segment for heavy and light chains. Native indicates the germline sequence identity of the original SAbDab sequence.}
    \label{fig:geneseqid}
\end{figure*}

In this appendix, we include a study of the germline distributions for the variable (V) and joining (J) gene segments.
For each of the 406 variable regions in the SAbDab test set, we sample 20 sequences with AbMPNN using $T=0.1$ and assign the closest germline match for each chain using ANARCI~\cite{anarci}.
Note that here we predict the full variable region, including the framework residues.
In Figures~\ref{fig:vgenedist} and~\ref{fig:jgenedist} we observe that the AbMPNN model closely follows the observed distribution of germline assignments, while with the original ProteinMPNN model applied to the same structures, ANARCI fails to annotate about 12.5\% of heavy chains and 9.4\% of light chains altogether.
In Figure~\ref{fig:geneseqid} we show the sequence identity with the closest germline, showing large improvement over ProteinMPNN with further fine-tuning on antibody-specific datasets. A summary of matching germline assignments for each model is shown in table~\ref{tab:germ}.

\begin{table}[]
\centering
\begin{tabular}{llcc}
\toprule
Model & Chain & V-gene & J-gene \\ 
\midrule
AbMPNN      & heavy & 0.56 & 0.59 \\
            & light & 0.55 & 0.62 \\ \addlinespace[0.5ex] 
SAbDab only & heavy & 0.51 & 0.58 \\
            & light & 0.52 & 0.63 \\ \addlinespace[0.5ex] 
ProteinMPNN & heavy & 0.44 & 0.42 \\
            & light & 0.48 & 0.45 \\ \addlinespace[0.5ex] 
OAS only    & heavy & 0.38 & 0.38 \\
            & light & 0.43 & 0.40 \\
\bottomrule
\end{tabular}
\caption{Fraction of matching V and J genes for each model.}
\label{tab:germ}
\end{table}



\end{document}